\documentclass[%
 reprint,
 amsmath,amssymb,
 aps,
]{revtex4-2}

\usepackage{graphicx}
\usepackage{dcolumn}
\usepackage{bm}
\usepackage{xcolor}

\newcommand{\tr}{{\rm tr }}
\newcommand{\p}{\partial}

\begin{document}

\preprint{APS/123-QED}

\title{Critical Lin-Lunin-Maldacena geometries}

\author{Prokopii Anempodistov}
 \email{anempodistov.pa@gmail.com}
\author{Vladimir Kazakov}%
 \email{vladimir.kazakov@phys.ens.psl.eu}

\author{Lev Senchukov}
\email{senchukov.lev@gmail.com}
 \altaffiliation[Also at]{ Institut Polytechnique de Paris, Palaiseau, France.}
 
\affiliation{%
 Laboratoire de Physique de l’École Normale Supérieure, CNRS, Université PSL, Sorbonne Universités, 
24 rue Lhomond, 75005 Paris, France
}%





\date{\today}

\begin{abstract}
We study the critical behavior of the Lin-Lunin-Maldacena (LLM) geometry in the case  when a droplet in the LLM base space develops a cusp. This cusp  is a generic feature of the density of complex eigenvalues in the dual complex matrix model (CMM) computing the correlation functions of huge  1/2-BPS operators in $\mathcal{N}=4$ SYM theory. It is also related to the criticality in CMM describing the pure $2D$ quantum gravity behavior.   The  supergravity dual -- LLM metric in the vicinity of the tip of the cusp -- acquires a universal $ISO(1,3)\times SO(5)$ symmetric form, with a naked singularity along a half-infinite line. Both massless and massive particles get trapped by this line singularity for almost any impact parameter. Generic trajectories ending on the singular line reach it in finite affine time, while the corresponding observer time diverges. An explicit analytic solution for a large class of massless trajectories together with the absence of stochastic behavior in the vicinity of the cusp  hint on a certain integrability of the problem.  
\end{abstract}

\maketitle
\raggedbottom

\section{Introduction}

Superstring theory is so far a unique theory providing stable (super)gravity vacua, staging a ground for a self-consistent fundamental physics. Among numerous examples of such vacua, those created by a stack of D3-branes in IIB type string theory, conserving a half of its supersymmetries,  played a special role for CFT/string dualities giving rise to the AdS/CFT  correspondence between light operators in $\mathcal{N}=4$ SYM and the string states on the $AdS_{5}\times S^{5}$ gravitational background~\cite{Maldacena:1997re,Gubser:1998bc,Witten:1998qj}. This duality, helped by quantum integrability~\cite{Beisert:2010jr,Gromov:2013pga,Gromov:2014caa,Kazakov:2018ugh,Gromov:2017blm,Basso:2013vsa,Basso:2015zoa},
 led to the complete (for any coupling) solution of the planar (large $N$) spectral problem and advanced computations of various correlators in  $\mathcal{N}=4$ SYM.  

But this duality is believed to hold even for finite $N$ and for any operators on the CFT side. In particular, "huge" operators ( $O \sim \exp{(C\,N^{2})}$) preserving a half of SUSY ($\frac12$-BPS) in  $\mathcal{N}=4$ SYM are dual to a specific background in Type IIB supergravity described in the classical limit \footnote{$N\gg\lambda\gg 1$, where $\lambda$  is the 't~Hooft coupling} by the LLM metric ~\cite{Lin:2004nb} - the most general non-singular $\frac12$-BPS background with $R \times SO(4)\times SO(4)$ symmetry. 
The striking property of the LLM background is that it is entirely determined by the shape of a  droplet in the 2d base plane which is identified with a phase space of $N$ free fermions. 
In the dual picture on the SYM side the two-point function of such  $\frac12$-BPS operators can be described by a complex matrix model (CMM) where the density of complex eigenvalues is constant on a domain whose shape exactly reproduces that of the dual LLM droplet~\cite{Takayama:2005yq,Vazquez:2006id, Berenstein:2004kk, Berenstein:2020jen, Skenderis:2007yb, Anempodistov:2025maj,Ghodsi:2005ks,Mosaffa:2006qk}.
For example, the SYM vacuum corresponds  to the disc shape of the LLM droplet rendering the $AdS_5 \times S^5$ background, and  Schur polynomial operators \cite{Corley:2001zk} in SYM correspond to the droplet given by concentric annuli.  
More general types of operators, such as exponential or coherent state ones~\cite{Vazquez:2006id, Berenstein:2022srd, Anempodistov:2025maj}, lead to virtually arbitrary shapes. In the vicinity of any smooth segment of the boundary of a droplet the LLM metric becomes of explicitly BMN type~\cite{Berenstein:2002jq,Lin:2004nb}. By deforming the parameters of an exponential operator, generically one can reach a situation when the droplet develops a cusp singularity known in relation to the Laplacian growth application of CMM~\cite{Zabrodin:2002up,Zabrodin:2004cc,Kazakov:2002yh,Kostov:2000ed,Teodorescu:2004qm}.

In this paper, we study the behavior of the LLM metric near this typical cusp singularity and its multi-critical generalizations for cusps with the 2d shape given by $x_{2}\sim x_{1}^{p/q},\quad p>q\in \mathbb{Z}_{+}$.   To study the vicinity of the cusp, we rescale appropriately the LLM coordinates and  establish a new  10d cusp metric with enhanced, $ISO(1,3)\times SO(5)$ symmetry, having a naked singularity along a half line starting from the tip of the cusp. This new universal cusp metric does not depend on the details of the bulk of the droplet for $1<p/q<3$. Then we study the behavior of massless and massive particles in this metric.    The trajectories of particles are rather regular  and for massless particles with zero $SO(5)$ charge we managed to solve  the equations of motion explicitly. These may be the signs of a certain classical integrability, though a real test for integrability would be for  a string $\sigma$-model in this background. The particle trapping phenomenon with a finite affine time but infinite observer time resembles the black hole behavior, though we deal here with a naked singularity.  Our cusp metric sets the stage for the study of the behavior of correlation functions of $\mathcal{N}=4$ SYM  near universal criticalities. 
We also study numerically the behavior of particles in the  full cusped LLM droplet of particular, "wing of the airplane" shape. For the blunt cusp  the typical classical trajectories of particles show in general a chaotic behavior, similarly to~\cite{Berenstein:2023vtd,Berenstein:2025ese}. For the sharp cusp the trajectories become much more ordered and seem to always end up on the tip  of the cusp.

\section{Cusp singularity for $\mathcal{N}=4$ SYM two-point correlators}

The protected  two-point correlation function of two local $\frac12$-BPS operators in $\mathcal{N}=4$ SYM can be cast into the complex matrix model~\cite{Vazquez:2006id,Anempodistov:2025maj}:
\begin{align}\label{corr}
    {G}(X_{1}-X_{2})&=\langle \bar{\mathcal{O}}(X_{1})\,\mathcal{O}_{2}(X_{2}) \rangle=
    \notag\\
    &=\int d\bar{Z}dZ e^{-N\tr \bar{Z}\,Z } \bar{O}(\bar{Z})O(Z),
\end{align}
where $Z$ is a $N\times N$ complex matrix and $\bar{Z}=Z^{\dagger}$. The operators we study   are of the exponential type:
\begin{align}
    O(Z)=\exp\left(-N\tr W(Z)\right),
\end{align}
where the holomorphic "potential" $W(z)=\sum_{k>0} g_{k} z^{k}$  can be rather arbitrary but must be able to confine, in the large $N$ limit, the complex eigenvalues of  $Z$  in a certain finite domain of the complex plane, as explained below. The distance $|X|=|X_{1}-X_{2}|$ dependence of the correlator can be absorbed into the trivial rescaling of the couplings of the operator $g_{n}\to |X|^{-n}g_{n}$, as explained in~\cite{Anempodistov:2025maj}.

Using the Schur decomposition $Z=U^{\dagger}(z+T)U$, where $U\in U(N)$, $T$ is  upper-triangular and $z=\text{diag}\{z_{1},z_{2},\dots,z_{N}\}$ is a  diagonal matrix encoding the complex eigenvalues, we can reduce the computation of the correlator to the e.v. integral
\begin{align*}
  G=\int \prod_{k=1}^{N} d^{2}z_{k} e^{-N [\bar{z}_k\,z_{k} +W(z_{k})+\bar{W}(\bar{z}_{k})] }|\Delta(z)|^{2},
\end{align*}
where $\Delta(z)=\prod_{k<j} (z_j-z_k)$ is the Vandermonde determinant.
Introducing the e.v. density $\rho(\bar{z},z)$ we write the saddle point equation in the large $N$ limit~\cite{Zabrodin:2004cc,Anempodistov:2025maj}
\begin{align}\label{saddle}
(\p_{\bar{z}}\p_{z})^{-1}\rho(\bar{z},z)= \bar{z}\,z +W(z)+\bar{W}(\bar{z})   .
\end{align}
Acting on both sides by 2d Laplacian $\p_{\bar{z}}\p_{z}$ we see that $\rho(\bar{z},z)=1/\pi,\,\, z\in \mathcal{D}$ and $\rho(\bar{z},z)=0,\,\, z\notin \mathcal{D}$, where $\mathcal{D}\in \mathbb{C}_{z}$ is a domain in the complex plane whose shape is defined by the holomorphic potential and the moduli of  the algebraic curve~\cite{Zabrodin:2004cc,Teodorescu:2004qm,Kostov:2000ed}.

As an example, we specify the potential as $W(z)=\frac{\tau}{2}z^{2}+\frac{1}{4}z^{4}$. Then the solution of~\eqref{saddle} is given by the algebraic curve projecting the interior of the unit circle $w=e^{i\phi},\,\,\phi\in (-\pi,\pi) $ into the droplet $\mathcal{D}$ of the "astroid" shape with the boundary given by $z(e^{i\phi})$~(see~\cite{Anempodistov:2025maj} and Fig.3 there)
\begin{equation}\label{AlgC}
    z(w) = r w - \frac{r \tau}{(3r^{2}-1)w}+\frac{r^{3}}{w^{3}},
\end{equation}
where the conformal radius $r$ is related to the area of the droplet as $\int_{\mathcal{D}}d^{2}z\equiv A=\pi r^2 \left(-3 r^4-\frac{\tau ^2}{\left(1-3 r^2\right)^2}+1\right)$.

For $r<r_{c}=\sqrt{\frac{1-\sqrt\tau}{3}}$ the contour is smooth, while at $r=r_{c}$
it develops  cusps $x_{2}-x_{2}^*\sim (x_{1}-x_{1}^*)^{3/2}$ at certain $(x_{1}^*,x_{2}^*)$  (here \(z=x_{1}+i\,x_{2}\)).  
The choice of the astroid shape is dictated by simplicity, but the results related to the cusp  singularity will not depend on the rest of the boundary shape. 

For example, the relation between the quantities  $\Delta=\int_{D}\frac{d^{2}z}{2\pi} \frac{\bar{z}z}{2}-\frac{1}{2}\left(\int_{D}\frac{d^{2}z}{2\pi})^{2}\right)^{2}$  and $A=4\pi^{2}N l_{p}^{4}$  representing in the dual theory the energy (dimension) of the state and the RR flux, resp.~\cite{Lin:2004nb},  near the critical point $A\to A_{0}$ of cusp formation is
\begin{align*}
    \Delta-\Delta_{0}=c_{1}\,(A_{0}-A)+ c_{3/2}\,(A_{0}-A)^{3/2}+\, O\left((A_{0}-A)^{2}\right).
\end{align*}
This type of singularity  is independent of the details  of the potential developing this cusp~\footnote{See in~\cite{Anempodistov:2025maj} the example of cubic  potentials giving the hypotrochoid  algebraic curve}.
This universal singular behavior gave rise to the matrix models of 2d quantum gravity and non-critical strings, starting from~\cite{Kazakov:1985ds,David:1984tx,Kazakov:1985ea}.  Similarly, the LLM metric generated by cusped droplets will be universal in the vicinity of a cusp and independent of the rest of the shape of the droplet~\footnote{This singularity (in that case $\sim \delta^{3/2}$) will persist in the protected three point correlator of two such huge $\frac12$-BPS operators and a sufficiently light probe (with dimension~$\ll N^2$).}.
     Moreover, similarly to the multicritical points of hermitian matrix models~\cite{Kazakov:1989bc,Daul:1993bg} we can also generalize and tune the CMM potential  in such a way that the droplet acquires a multicritical cusp of the shape $x_{2}\sim x_{1}^{p/q}$, where $p>q\in \mathbb{Z}_{+}$. E.g. for  astroid we have at $\tau=4$  the multicritical point $(p,q)=(4,3)$~\cite{Anempodistov:2025maj}.

     According to the duality discussed above, we will associate such droplets with the LLM geometries and study their critical behavior in the vicinity of such  cusps.   
     
\section{Universal LLM metric near the cusp}

The two-point correlation function described above has its dual gravity realization in terms of the LLM geometry given, for arbitrary shape of droplets, by the metric~\cite{Lin:2004nb}
\begin{align}\label{LLMmetric}
&ds^{2}=-\frac{y}{\sqrt{\frac{1}{4}-\zeta^{2}}}(dt-\frac{i}{2}\bar{V}dz+\frac{i}{2}Vd\bar{z})^{2}+\\
&
+\sqrt{\frac{1}{4}-\zeta^{2}}\frac{dy^{2}+d\bar{z}dz}{y}+y\sqrt{\frac{\frac{1}{2}+ \zeta}{\frac{1}{2}- \zeta}}d\Omega^{2}_{3}+y\sqrt{\frac{\frac{1}{2}- \zeta}{\frac{1}{2}+\zeta}}d\tilde\Omega_{3}^{2},   
\notag\end{align}
\begin{align}\label{Vdef}
\text{where}\,\,\, V(z,\bar z,y)
&=
\int_D \frac{d^2z'}{\pi}\,
\frac{z-z'}{\left(|z-z'|^2+y^2\right)^2},
\\
\zeta(z,\bar z,y)-\frac12&=
-\int_D \frac{d^2z'}{\pi}\,
\frac{y^2}{\left(|z-z'|^2+y^2\right)^2}\,.
\label{zetadef}\end{align}
and $\mathcal{D}$ is the interior of a droplet~\eqref{AlgC}.  $V(\bar{z},z,y)$ and $\zeta(\bar{z},z,y)$  can be computed for the astroid shaped droplet~\eqref{AlgC}   explicitly but instead  we prefer to work with a general cusped droplet and concentrate on the vicinity of the cusp  which is placed at $z\equiv x_{1}+ix_{2}=0,\,\,\,y=0$ and has the shape of the boundary $x_{2}=\pm|x_{1}|^{3/2}+O((|x_{1}|)^{5/2})$. Moreover, we generalize this statement to  multi-critical cusps $x_{2}\simeq(|x_{1}|)^{p/q}$ where $p,q\in\mathbb{Z}_{+}$  and $\alpha=p/q$.     We will show, that for $1<\alpha<3$ the LLM metric and the flux field strength near the cusp is universal and depends only on $\alpha$: it is completely defined by the type of the cusp and does not depend on the shape of the rest of the droplet.   
To show it we  rescale the LLM coordinates as follows \footnote{Notice that in this scaling the metric sees only the flat tangent space $\mathbb{R}^{3}$ of the sphere $\tilde{S}^{3}$.}:
\begin{align*}
 x_{1,2}\to \epsilon\, x_{1,2},\quad  y\to \epsilon\, y, 
 \quad t\to \epsilon^{\frac{\alpha-1}{2}}t,\quad d\tilde{\Omega}_{3}\to \epsilon^{\frac{\alpha-1}{2}} d\mathbb{R}_3
\end{align*}
and separate in the integrals~\eqref{Vdef}-\eqref{zetadef} two domains: one is around the cusp $0<|x_{1}|<\epsilon$ and another one is the rest of the droplet $x_{1}>\epsilon$ \footnote{We assume for simplicity that the bulk of the droplet is smooth and resides entirely to the left of the cusp, as in Fig.\ref{fig:chaos}}.
The rescaled metric $ds^2_{\rm crit}= \epsilon^{-\frac{\alpha+1}{2}}\, ds^2$  takes a simple form in the limit $\epsilon\to 0$ 
\begin{align}\label{cuspmetric}
&ds_{\mathrm{crit}}^2
=
\mathcal {H}_\alpha(r,\theta)\,
ds_{\mathbb R^{1,3}}^2
+\mathcal{H}_\alpha(r,\theta)^{-1} \times \notag\\
&\times
\left(
dr^2+r^2d\theta^2+r^2\sin^2\theta\,d\Omega_4^2
\right)+O\left(\epsilon^{\beta}\right),
\end{align}
with $\beta \equiv \frac12\min(\alpha-1,3-\alpha)$ and the universal critical warp factor: 
\begin{align*}
\mathcal H_\alpha(r,\theta)=r^{\frac{3-\alpha}{2}}\left[\frac{
-2\sin^{3}\theta\,\sin(\pi\alpha)}{\left((\alpha-2)\sin(\alpha\theta)
+\alpha\sin((2-\alpha)\theta)\right)}
\right]^{1/2},
\end{align*}
where we used the spherical coordinates 
$x_1=r\cos\theta
$, $
x_2=r\sin\theta\cos\varphi$,
 $
y=r\sin\theta\sin\varphi $.  The details of derivation can be found in Supplementary Material. 
 This cusp metric obeys an extended, with respect to the general LLM metric~\eqref{LLMmetric}, symmetry $ISO(1,3)\times SO(5)$. 
 
 One can even consider a "blunt" cusp with the boundary parameterized in the vicinity of the cusp as 
$z(\varphi)=-\varphi^q+i\varphi^p+i\delta\varphi+o(\varphi^p)$ with appropriate scaling of the "bluntness" parameter $\delta\mapsto\epsilon^\frac{p-1}{q}\delta$. Our metric then  generalizes to the same form~\eqref{cuspmetric} but   with a universal modified warp factor:
 \begin{align*}
\frac{1}{\mathcal{H}^2_{\alpha,\delta}}=\frac{1}{\mathcal{H}^2_{\alpha}}+\frac{\delta}{\mathcal{H}^2_{\frac{1}{q}}}\,.
\end{align*}
In fact, the same  metric \eqref{cuspmetric}  can be derived by considering a stack of D3-branes   distributed on a cusp described above (see Supplementary Material for the details of the derivation). Moreover, the function $\mathcal{H}^{-2}_{\alpha,\delta}(r,\theta)$ is harmonic in the 6d space, which means that this brane configuration is dual to a particular moduli distribution of the Coulomb branch of $\mathcal{N}=4$ SYM~\footnote{We thank Shota Komatsu and Juan Maldacena for bringing this important point to our attention.}.  Notice that the rotational symmetry around the $x_1$ axis completes the LLM 3-sphere $S^3$  to the  4-sphere $S^4$ and 3d tangent space  of the $R$ symmetry sphere $\tilde{S}^3$ forms, together with the time, the Minkowski (3+1)d flat space. 
 Notice also that the metric~\eqref{cuspmetric} scales as a non-integer power of $\epsilon$ which makes it distinguishable from large but regular in $\epsilon$ contributions from the bulk of the droplet.
 
 The leading five-form flux
is also universal in this limit: 
 \begin{align*}
F_{5,\rm crit}
=
-\frac14\, d(H_\alpha^2)\wedge dt\wedge d^3\widetilde x
+
\frac14\, y^3 *_3 d(H_\alpha^{-2})\wedge d\Omega_3 .
\end{align*}

\section{ Naked singularity }

The LLM metric for arbitrary droplets with smooth boundaries is known to be non-singular everywhere in 10d space. Our cusp metric is singular due to the singular shape of the boundary at the cusp. Although the Ricci scalar $\mathcal{R} = 0$ as for the generic LLM metric,
we detect the singularity using the Kretschmann scalar $\mathcal{K}$. Its explicit form is available  for general $\alpha$  but it is lengthy. For the  simplest and most important $\alpha=\frac32 $ case it is
\begin{align*}
    \mathcal{K}&\equiv R_{MNPQ}R^{MNPQ}
=\frac{9  \left(32 \cos^{2} \! \left(\frac{\theta}{2}\right)+73 \right)}{32 r^{{5}/{2}}\cos \! \left(\frac{\theta}{2}\right)}.
\end{align*}
The singularity (for any $\alpha)$ occurs when we approach the  half-axis $x_{1}=\Re (z)<0$ from above or from below:  $\mathcal{K}
\sim\frac{657}{16}r^{-(\alpha+1)}
(\pi\mp\theta)^{-1}$ for $\theta\to\pm\pi$. The  singularity at the tip of the cusp is enforced by an $\alpha$-dependent power of the distance to the cusp.

 So our cusp metric has a naked singularity sourced by an open half-line~$x_{1}\in (-\infty,0]\cap x_{2}=0$.
\section{Cusp metric traps   particles  }

We will show here that the behavior of classical massless and massive particles in the cusped metric is drastically different from the one in generic LLM metric with smooth boundaries of droplets. 

First we  consider   the most representative cusp \((p,q)=(3,2)\) for which ~\eqref{cuspmetric} gives
\begin{align*}
&ds_{\rm crit}^2
=
2\sqrt{2}r^{3/4}\cos^{3/2}\frac{\theta}{2}
\left(
-dt^2+d\mathbf{s}_{\mathbb{R}^{3}}^2
\right)+
\\&+\frac{1}{2\sqrt{2}r^{3/4}\cos^{3/2}\frac{\theta}{2}}
\left(
dr^2+r^2 d\theta^{2}+r^2\sin^{2}\theta d\Omega_4^2
\right).
\end{align*}

For a massless particle, we can write the following reduced action:
\begin{align*}
S_{\rm red}
=
\int dt
\left(
p_r \dot r + p_\theta \dot\theta - E
\right).
\end{align*}
where the energy $E$ and conserved charges $\vec{P}$ and $J_{4}$ related to the two isometries are constrained by
\begin{align*}
E^2
=
\vec P^2
+
8r^{3/2}\cos^3\frac{\theta}{2} p_r^2
+
8r^{-1/2}\cos^3\frac{\theta}{2}
\left(
p_\theta^2+\frac{J_4^2}{\sin^2\theta}
\right).
\end{align*}
For the already quite a representative case $J_{4}=0$ the motion is confined to a 2d plane and all the trajectories $r(\theta)$,  except for the degenerate cardioid case given by $r(\theta)=r_0\sin^2\frac{\theta}{2}$, can be obtained in explicit parametric form.    Namely, we show in the Appendix A that the observed trajectory is given explicitly as a \(\xi\)-parameterization
\begin{align}
    \label{R(xi)}
\tan\frac{\theta(\xi)}{2}
&=
\xi^{-1/3}
\left[
C_\theta
-
\frac{\sigma_2}{3}
B_\xi\!\left(\frac56,\frac12\right)
\right],\notag\\
r(\xi)
&=
\frac{C_R}{
\xi^{2/3}
+
\left[
C_\theta
-
\frac{\sigma_2}{3}
B_\xi\!\left(\frac56,\frac12\right)
\right]^2
}.
\end{align}
where $B_{\xi}(\alpha,\beta)$ is the incomplete beta function and $\sigma_i=\pm1$ is defined by the initial conditions (we provide more details in the appendix). The observer time is
\begin{align*}
&t(\xi)=
-\sigma_1\sigma_2\frac{E\,C_R^{1/4}}{3\sqrt{2}P}
\left[
B_\xi\!\left(-\frac16,\frac12\right)
-
B_{\xi_0}\!\left(-\frac16,\frac12\right)
\right].
\end{align*}

The two constants fix the boundary conditions:
\begin{align*}
C_\theta
&=
\xi_0^{1/3}
\tan\frac{\theta_0}{2}
+
\frac{\sigma_2}{3}
B_{\xi_0}
\left(
\frac56,\frac12
\right),
\qquad C_R=
\frac{r_0
\xi_0^{\frac23}}{\cos^2\frac{\theta_0}{2}}.
\end{align*}

\begin{figure}[t]
    \centering
    \includegraphics[width=0.8\columnwidth]{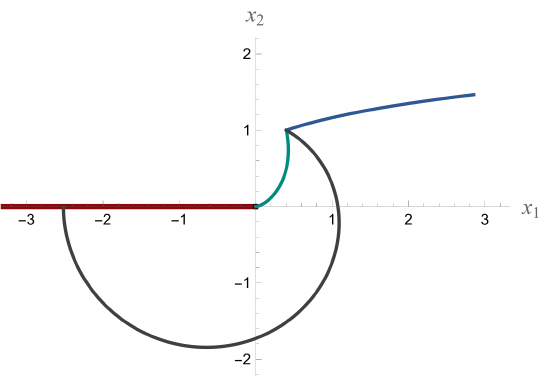}
    \caption{
Trajectories of massless particles: if sent from the same point with different initial angles, almost all of them, apart from the separatrix escaping to infinity (seen on the right of the picture) end up at the singularity line.   } 
    \label{fig:massless}
\end{figure}

We  find from this  solution  
 that our cusp metric  behaves as an attractor and it traps massless  particles for almost all initial conditions, apart from a special separatrix corresponding to $C_{\theta}=0$ that can escape arbitrarily far away from the singular half-line. The corresponding solutions can be continued to arbitrarily large values of both the affine parameter and the observer time. The trapped massless particles can follow two different types of trajectories. The first ones are  generic trajectories: they reach the singular cut during a finite interval of affine parameter, but during an infinite observer time. The second ones are the cardioid trajectories (including the one  moving along the $x$-axis). These trajectories end up at the origin and reach it in both finite affine parameter and finite observer time~(see Fig.\ref{fig:massless}). 

\subsection{Massive particle}

For a massive particle moving along a timelike curve $\gamma(u)$, $u\in[0,1]$
the action is
\begin{align*}
S[\gamma]
=
m\int_0^1 du\,
\sqrt{
-g_{\mu\nu}\bigl(\gamma(u)\bigr)\,
\frac{d\gamma^\mu}{du}\,
\frac{d\gamma^\nu}{du}
}.
\end{align*}

The critical metric admits the scaling transformation
$ds_{crit}\to \lambda^{\frac{\alpha+1}{4}} ds_{crit}$ when $\theta\mapsto\theta,
\Omega_4\mapsto\Omega_4$ and
\begin{align*}
r\mapsto \lambda r,&
\qquad t
\mapsto
\lambda^{\frac{\alpha-1}{2}} t,\qquad
\tilde{x}^i\mapsto \lambda^{\frac{\alpha-1}{2}}\tilde{x}^i.
\end{align*}
Under this transformation the conserved charges scale as:
\begin{align*}
E=\mathcal H_\alpha\dot t\mapsto\lambda^{\frac{3-\alpha}{4}}E, \quad\vec{P}\mapsto\lambda^{\frac{3-\alpha}{4}}\vec{P},\quad J_4\mapsto\lambda^{\frac{\alpha+1}{4}}J_4.
\end{align*}
And the action scales as:
\begin{align*}
S[\gamma]\, \mapsto\lambda^{\frac{\alpha+1}{4}}S[\gamma].
\end{align*}
This allows us to conclude that 
\begin{align*}
S=CE^{\frac{\alpha+1}{3-\alpha}}=C'(\Delta t)^{\frac{\alpha+1}{2(\alpha-1)}}.
\end{align*}
where $C$, $C'$ depend only on dimensionless initial conditions.

The trajectories $r(\theta)$ of massive particles at $J_{4}=0$ for arbitrary  $\alpha=p/q$ are subject to the 2nd order ODE:
\begin{align*}
&r''=r+2\frac{r'^2}{r}+
\frac{r'^2+r^2}{r}
\left(
1+\frac{\mathcal H_\alpha}{2(P^2-\mathcal H_\alpha)}
\right)\times\\&\times
\left(
\frac{r'}{r}\,\partial_\theta\log \mathcal H_\alpha
-
\frac{3-\alpha}{2}
\right).
\end{align*}
Our analysis  suggests that all massive-particle trajectories are confined and do not escape to infinity. If a trajectory ends up on the singular cut, the corresponding observer time diverges. In contrast, trajectories that reach the cusp tip do so in finite observer time.
\begin{figure}[t]
    \centering
\includegraphics[width=0.7\columnwidth]{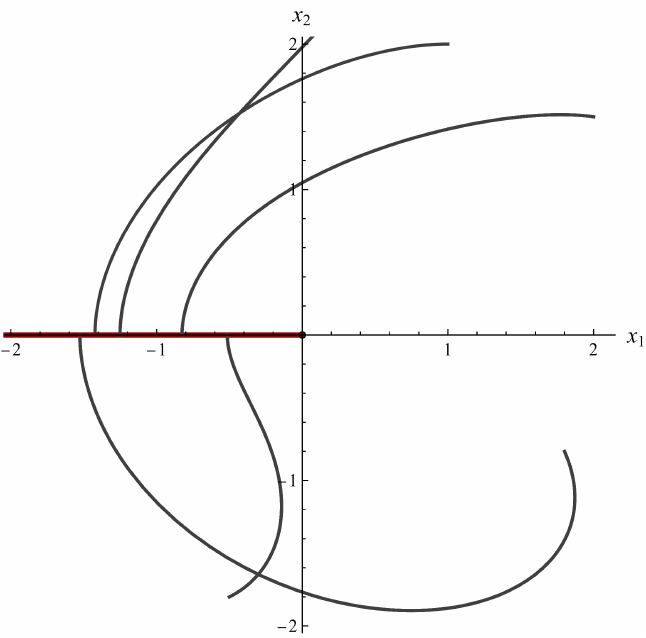}
    \caption{
        All trajectories of massive particles built numerically 
        are trapped by the singularity line, from above or from below, or  at the tip, depending on the initial conditions. } 
    \label{fig:massive}
\end{figure}

\begin{figure}[t] 
    \centering
\includegraphics[width=1.0\columnwidth]{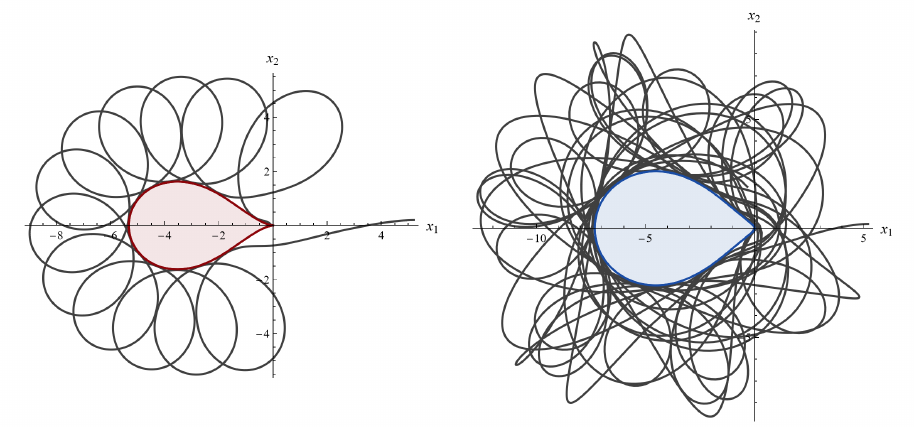}
    \caption{
        The left picture shows a typical massless trajectory in the sharp-cusp wing metric, which looks like a cycloid turning around the wing until it ends up on the cusp. On the right picture, there is a typical chaotic trajectory for a wing with a blunt cusp. } 
    \label{fig:chaos}
\end{figure}

It is also instructive to study the trajectories in the metric~\eqref{LLMmetric} induced by the full  droplet~\eqref{AlgC} which we choose here for simplicity of the "airplane wing" shape $z(w)= 
r w+u +\frac{v}{w-a}$. The  typical trajectories in the $y=0$ plane for wing droplets with a sharp and blunt cusps are presented in  Fig.\ref{fig:chaos}. 
In the vicinity of a blunt cusp, the typical trajectory becomes chaotic, similarly to the observations of~\cite{Berenstein:2023vtd,Berenstein:2025ese}. For the sharp cusp, they suddenly become more regular, looking like cycloids winding around the wing and eventually ending up at the cusp.

\section{Discussion}

In this work we obtained a universal LLM metric in the vicinity of a cusp of a droplet, having a naked singularity along a half line in 10d space. We stress that this cusp is a generic feature for any $\frac12$-BPS state/operator in string/SYM duality: changing the parameters of any operator, i.e. evolving the LLM droplet, we inevitably hit such a cusp singularity, except for the case of elliptic or  radially symmetric droplets. The massless and massive particles are trapped in this metric at the naked singularity situated along a half infinite line, with a typical slowing down in observer time (infinite redshift) which resembles the behavior of particles near a black-hole horizon.   It would be interesting to use this cusp metric to compute  protected 3-point correlators involving   2 huge operators with cusp singularity  and one giant graviton and to compare them to the exact $\mathcal{N}=4$ SYM predictions which can be  computed using the matrix techniques of~\cite{Kazakov:2024ald} and the methods of~\cite{Anempodistov:2026dhi}, or even non-protected 4-point correlators as in~\cite{Turton:2024afd,Aprile:2025hlt}.  The singularity of the metric should produce some universal singularities for such correlators. The fact that the trajectories in our cusp metric become quite regular and that the massless-particle equations can be solved in quadrature hints on a possible integrability of the problem. Of course the real test of integrability is the string $\sigma$-model in this metric or, on the SYM side, integrability of the quantum spin chain  interacting  with such a cusp in the CMM. It would be also interesting to understand the quantization of the LLM gravity near the cusp and find the string dual to the universal double scaling of the SYM matrix model in analogy with~\cite{Brezin:1990rb,Douglas:1989ve,Gross:1989vs,Douglas:1989dd}. Another curious fact about our cusp metric: it contains the flat space $\mathbb{R}^{(1,3)}$ at any $r,\theta$. Does it define a new 4d SCFT?
Moreover, one could try to find the analogs of the cusp metric for the bubbling 1/4- and 1/8-BPS geometries~\cite{Donos:2006iy,Chen:2007du,Gava:2006pu,Lunin:2008tf}. Finally, let us point out an intriguing feature of the scaling limit. The metric near the cusp coincides exactly with the local geometry sourced by a stack of D3-branes distributed along a cuspidal domain in the transverse directions. In the dual gauge theory, this means that the scaling limit isolates a Coulomb branch configuration of SYM \cite{Freedman:1999gk,Brandhuber:1999jr,Skenderis:2006di,Skenderis:2006uy}, with a cusp-like distribution of scalar eigenvalues. It would be interesting to understand how this Coulomb branch dynamics arises from the original description, and whether this fact can be used to simplify the calculation of heavy three-point functions.

\vspace{0.5cm}
\begin{acknowledgments}
We thank A.~Holguin, S.~Komatsu, I.~Kostov, O.~Lunin, J.~Maldacena, H.~Murali, P.~Vieira  and K.~Zarembo  for useful discussions and remarks on the manuscript. We are especially grateful to S.~Komatsu and J.~Maldacena for important comments and proposals.
\end{acknowledgments}

\appendix
\label{app:particles&strings}
\section{Particle  in critical LLM metric}

We first record the reduction of the geodesic equations. We denote differentiation with respect to an affine parameter \(s\) by a dot.
The translational isometries in \(\mathbb R^{1,3}\) give
\begin{align*}
E=\mathcal H_\alpha \dot t,
\qquad
P_i=\mathcal H_\alpha \dot{\tilde{x}}_i,
\qquad
P^2=E^2-\vec P^{\,2}.
\end{align*}
By the \(SO(5)\) symmetry, the motion on \(S^4\) can be restricted to a great
circle, with conserved angular momentum
$J_4=\mathcal H_\alpha^{-1}r^2\sin^2\theta\,\dot\varphi$ .
Eliminating the cyclic variables and writing an effective action for others, one obtains:
\begin{align*}
S_{\rm red}
=
\frac12\int ds\,
\left[
\mathcal H_\alpha^{-1}
\left(
\dot r^{\,2}+r^2\dot\theta^{\,2}+P^2
\right)
-
\mathcal H_\alpha
\frac{J_4^2}{r^2\sin^2\theta}
\right].
\end{align*}

In what follows we restrict to the sector of vanishing angular momentum on
\(S^4\),
$J_4=0$ . The normalization condition is
$
g_{\mu\nu}\dot X^\mu\dot X^\nu=\sigma
$, with $\sigma=0$ for massless and $\sigma=-1$ for massive-particles trajectories and therefore
\begin{align*}
\dot r^{\,2}+r^2\dot\theta^{\,2}
=
P^2+\sigma\mathcal H_\alpha(r,\theta).
\end{align*}

Firstly, let us consider null geodesics with \(J_4=0\) and $\alpha=\frac32$. 
The Euler-Lagrange equations for null geodesics are
\begin{align*}
\ddot r-r\dot\theta^{\,2}
-
\left(
\frac{3\dot r}{4r}
-\frac34\tan\frac{\theta}{2}\,\dot\theta
\right)\dot r
+
\frac{3P^2}{4r}
=0,
\end{align*}
\begin{align*}
r^2\ddot\theta+2r\dot r\dot\theta
-
\left(
\frac{3\dot r}{4r}
-\frac34\tan\frac{\theta}{2}\,\dot\theta
\right)r^2\dot\theta
-
\frac{3P^2}{4}\tan\frac{\theta}{2}
=0,
\end{align*}

Introduce the angle $\varphi$  defined by
\begin{align*}
\sin\varphi=
\frac{
r\dot\theta\cos\frac{\theta}{2}
-\dot r\sin\frac{\theta}{2}
}{P},\qquad\xi=\sin^2\varphi,\qquad \eta=\tan\frac{\theta}{2}
\end{align*}
\begin{align*}
\sigma_1=\mathrm{sgn}(\sin\varphi_0),\qquad \sigma_2=\mathrm{sgn}(\sin2\varphi_0)\,.
\end{align*}
The above parametrization is valid on each monotonic branch \(0<\xi<1\). If a trajectory reaches \(\xi=1\), it can be smoothly continued into a new branch, with the integration constants fixed by matching \(r\), \(\theta\), and \(t\) at \(\xi=1\).

Differentiating \(\xi\) and $\eta$ w.r.t. affine parameter and using the equations of motion we have
\begin{align*}
\dot\xi
&=
-\sigma_1\sigma_2
\frac{3P}{2r\cos\frac{\theta}{2}}\,
\xi\sqrt{1-\xi},
\\
\dot \eta&=
\sigma_1\frac{P}{2r\cos\frac{\theta}{2}}
\left(
\sqrt{\xi}
+\sigma_2 \eta\sqrt{1-\xi}
\right).
\end{align*}
In the regular region
\begin{equation*}
    \dot\xi=0
\qquad\Longleftrightarrow\qquad
\xi=0
\quad\text{or}\quad
\xi=1.
\end{equation*}
$\xi=1$ is not a solution of the equations of motion. On the other hand, \(\xi=0\) provides us with the cardioid solution $r(\theta)=r_0\sin^2\frac{\theta}{2}$.
For a non-cardioid geodesic, one may divide $\dot \eta$ by $\dot\xi$. This gives
 the linear equation
 \begin{align*}
\frac{d\eta}{d\xi}
+
\frac{\eta}{3\xi}
&=
-\frac{\sigma_2}{3\sqrt{\xi(1-\xi)}},\\
\frac{d}{d\xi}\bigl(\xi^{1/3}\eta\bigr)
&=
-\frac{\sigma_2}{3}\,
\xi^{-1/6}(1-\xi)^{-1/2}.
\\
\text{with solution}\quad\eta(\xi)
&=
\xi^{-1/3}
\left[
C_\theta
-
\frac{\sigma_2}{3}\,
B_\xi\!\left(\frac56,\frac12\right)
\right],
\\
C_\theta
&=
\xi_0^{1/3}
\tan\frac{\theta_0}{2}
+
\frac{\sigma_2}{3}
B_{\xi_0}
\left(
\frac56,\frac12
\right).
\end{align*}
Restoring $r(\xi)$ from the introduction of $\xi$ one has:
\begin{align*}
&r(\xi)
=
\frac{C_R}{\xi^{2/3}\bigl(1+\eta(\xi)^2\bigr)}=
\frac{C_R}{
\xi^{2/3}
+
\left[
C_\theta
-
\frac{\sigma_2}{3}
B_\xi\!\left(\frac56,\frac12\right)
\right]^2
},\\& C_R=
\frac{r_0
\xi_0^{\frac23}}{\cos^2\frac{\theta_0}{2}}.
\end{align*}
And for $t$:
\begin{align*}
\frac{dt}{d\xi}
=
\frac{\dot t}{\dot\xi}
=
\frac{E}{\mathcal H_{3/2}\dot\xi}
=
-\sigma_1\sigma_2\frac{E\,C_R^{1/4}}
{3\sqrt{2}P\,\xi^{7/6}\sqrt{1-\xi}}.
\end{align*}
The solution is
\begin{align*}
\tan\frac{\theta(\xi)}{2}
&=
\xi^{-1/3}
\left[
C_\theta
-
\frac{\sigma_2}{3}
B_\xi\!\left(\frac56,\frac12\right)
\right],
\\
r(\xi)
&=
\frac{C_R}{
\xi^{2/3}
+
\left[
C_\theta
-
\frac{\sigma_2}{3}
B_\xi\!\left(\frac56,\frac12\right)
\right]^2
},
\\
t(\xi)
&=
-\sigma_1\sigma_2\frac{E\,C_R^{1/4}}{3\sqrt{2}P}
\left[
B_\xi\!\left(-\frac16,\frac12\right)
-
B_{\xi_0}\!\left(-\frac16,\frac12\right)
\right].
\end{align*}
Since $t\to\infty$ corresponds to $\xi\to0$
\begin{align*}
\tan\frac{\theta}{2}
=
\begin{cases}
C_\theta\,\xi^{-1/3}
+O\!\left(\xi^{1/2}\right),
&
 C_\theta\neq0
 \\
-\dfrac{2\sigma_2}{5}\,\xi^{1/2}
+O\!\left(\xi^{3/2}\right),& C_\theta=0
\end{cases}
\end{align*}
\begin{align*}
r(\xi)=
\begin{cases}
\dfrac{C_R}{C_\theta^2}
+O\!\left(\xi^{2/3}\right),
& C_\theta\neq0
\\
C_R\,\xi^{-2/3}+
O\!\left(\xi^{1/3}\right),
& C_\theta=0
\end{cases}
\end{align*}
All trajectories apart from the special separatrix solution end up on the cut.

Notice that
we can consider a  consistent truncation $\Omega_4=\mathrm{const},
\quad
\theta=0$.
In the resulting metric the massive particles will move along the axis $x_{1}$ between the tip of the cusp and the turning point $r_{turn}=\frac14P^\frac83$.

\bibliography{Cusp}

@article{Kostov:2000ed,
    author = "Kostov, I. K. and Krichever, I. and Mineev-Weinstein, M. and Wiegmann, P. B. and Zabrodin, A.",
    title = "{Tau function for analytic curves}",
    booktitle = "{Introductory Workshop on Random Matrix Models and their Applications}",
    eprint = "hep-th/0005259",
    archivePrefix = "arXiv",
    month = "5",
    year = "2000",
    journal = {}
}

@article{Lin:2004nb,
    author = "Lin, Hai and Lunin, Oleg and Maldacena, Juan Martin",
    title = "{Bubbling AdS space and 1/2 BPS geometries}",
    eprint = "hep-th/0409174",
    archivePrefix = "arXiv",
    reportNumber = "PUPT-2136",
    doi = "10.1088/1126-6708/2004/10/025",
    journal = "JHEP",
    volume = "10",
    pages = "025",
    year = "2004"
}

@article{Kazakov:2002yh,
    author = "Kazakov, Vladimir A. and Marshakov, Andrei",
    title = "{Complex curve of the two matrix model and its tau function}",
    eprint = "hep-th/0211236",
    archivePrefix = "arXiv",
    reportNumber = "LPTENS-02-60, IHES-P-02-83, FIAN-TD-15-02, ITEP-TH-56-02",
    doi = "10.1088/0305-4470/36/12/315",
    journal = "J. Phys. A",
    volume = "36",
    pages = "3107--3136",
    year = "2003"
}

@article{Ghodsi:2005ks,
    author = "Ghodsi, A. and Mosaffa, A. E. and Saremi, O. and Sheikh-Jabbari, M. M.",
    title = "{LLL vs. LLM: Half BPS sector of N=4 SYM equals to quantum Hall system}",
    eprint = "hep-th/0505129",
    archivePrefix = "arXiv",
    reportNumber = "IPM-P-2005-034",
    doi = "10.1016/j.nuclphysb.2005.08.042",
    journal = "Nucl. Phys. B",
    volume = "729",
    pages = "467--491",
    year = "2005"
}

@article{Mosaffa:2006qk,
    author = "Mosaffa, A. E. and Sheikh-Jabbari, M. M.",
    title = "{On classification of the bubbling geometries}",
    eprint = "hep-th/0602270",
    archivePrefix = "arXiv",
    reportNumber = "IPM-P-2006-015",
    doi = "10.1088/1126-6708/2006/04/045",
    journal = "JHEP",
    volume = "04",
    pages = "045",
    year = "2006"
}

@article{Kazakov:2024ald,
    author = "Kazakov, Vladimir and Murali, Harish and Vieira, Pedro",
    title = "{Huge BPS Operators and Fluid Dynamics in $\mathcal{N}=4$ SYM}",
    eprint = "2406.01798",
    archivePrefix = "arXiv",
    primaryClass = "hep-th",
    month = "6",
    year = "2024",
    journal = {}
}

@article{Kazakov:1989bc,
    author = "Kazakov, V. A.",
    editor = "Brezin, E. and Wadia, S. R.",
    title = "{The Appearance of Matter Fields from Quantum Fluctuations of 2D Gravity}",
    reportNumber = "NBI-HE-89-25",
    doi = "10.1142/S0217732389002392",
    journal = "Mod. Phys. Lett. A",
    volume = "4",
    pages = "2125",
    year = "1989"
}

@article{Freedman:1999gk,
    author = "Freedman, D. Z. and Gubser, S. S. and Pilch, K. and Warner, N. P.",
    title = "{Continuous distributions of D3-branes and gauged supergravity}",
    eprint = "hep-th/9906194",
    archivePrefix = "arXiv",
    reportNumber = "CERN-TH-99-189, HUTP-99-A029, MIT-CTP-2877, USC-99-03",
    doi = "10.1088/1126-6708/2000/07/038",
    journal = "JHEP",
    volume = "07",
    pages = "038",
    year = "2000"
}

@article{Brandhuber:1999jr,
    author = "Brandhuber, A. and Sfetsos, K.",
    title = "{Wilson loops from multicenter and rotating branes, mass gaps and phase structure in gauge theories}",
    eprint = "hep-th/9906201",
    archivePrefix = "arXiv",
    reportNumber = "CERN-TH-99-191",
    doi = "10.4310/ATMP.1999.v3.n4.a4",
    journal = "Adv. Theor. Math. Phys.",
    volume = "3",
    pages = "851--887",
    year = "1999"
}

@article{Skenderis:2006di,
    author = "Skenderis, Kostas and Taylor, Marika",
    title = "{Holographic Coulomb branch vevs}",
    eprint = "hep-th/0604169",
    archivePrefix = "arXiv",
    reportNumber = "ITFA-2006-18",
    doi = "10.1088/1126-6708/2006/08/001",
    journal = "JHEP",
    volume = "08",
    pages = "001",
    year = "2006"
}

@article{Skenderis:2006uy,
    author = "Skenderis, Kostas and Taylor, Marika",
    title = "{Kaluza-Klein holography}",
    eprint = "hep-th/0603016",
    archivePrefix = "arXiv",
    reportNumber = "ITFA-2006-04",
    doi = "10.1088/1126-6708/2006/05/057",
    journal = "JHEP",
    volume = "05",
    pages = "057",
    year = "2006"
}

@article{Vazquez:2006id,
    author = "Vazquez, Samuel E.",
    title = "{Reconstructing 1/2 BPS Space-Time Metrics from Matrix Models and Spin Chains}",
    eprint = "hep-th/0612014",
    archivePrefix = "arXiv",
    doi = "10.1103/PhysRevD.75.125012",
    journal = "Phys. Rev. D",
    volume = "75",
    pages = "125012",
    year = "2007"
}

@article{Brezin:1990rb,
    author = "Brezin, E. and Kazakov, V. A.",
    title = "{Exactly Solvable Field Theories of Closed Strings}",
    doi = "10.1016/0370-2693(90)90818-Q",
    journal = "Phys. Lett. B",
    volume = "236",
    pages = "144--150",
    year = "1990"
}

@article{Teodorescu:2004qm,
    author = "Teodorescu, R. and Bettelheim, E. and Agam, O. and Zabrodin, A. and Wiegmann, P.",
    title = "{Normal random matrix ensemble as a growth problem: Evolution of the spectral curve}",
    eprint = "hep-th/0401165",
    archivePrefix = "arXiv",
    doi = "10.1016/j.nuclphysb.2004.10.006",
    journal = "Nucl. Phys. B",
    volume = "704",
    pages = "407--444",
    year = "2005"
}

@article{Zabrodin:2002up,
    author = "Zabrodin, A.",
    editor = "Rivasseau, V.",
    title = "{New applications of nonHermitian random matrices}",
    eprint = "cond-mat/0210331",
    archivePrefix = "arXiv",
    reportNumber = "ITEP-TH-46-02",
    doi = "10.1007/s00023-003-0966-2",
    journal = "Annales Henri Poincare",
    volume = "4",
    pages = "S851--S861",
    year = "2003"
}

@article{Douglas:1989dd,
    author = "Douglas, Michael R.",
    editor = "Brezin, E. and Wadia, S. R.",
    title = "{Strings in Less Than One-dimension and the Generalized $K^- D^- V$ Hierarchies}",
    reportNumber = "RU-89-51",
    doi = "10.1016/0370-2693(90)91716-O",
    journal = "Phys. Lett. B",
    volume = "238",
    pages = "176",
    year = "1990"
}

@article{Gross:1989vs,
    author = "Gross, David J. and Migdal, Alexander A.",
    editor = "Brezin, E. and Wadia, S. R.",
    title = "{Nonperturbative Two-Dimensional Quantum Gravity}",
    reportNumber = "PUPT-1148",
    doi = "10.1103/PhysRevLett.64.127",
    journal = "Phys. Rev. Lett.",
    volume = "64",
    pages = "127",
    year = "1990"
}

@article{Berenstein:2025ese,
    author = "Berenstein, David and {\v{C}}ubrovi{\'c}, Mihailo and Djuki{\'c}, Vladan",
    title = "{Trapping, chaos and averaging in bubbling AdS spaces}",
    eprint = "2508.09669",
    archivePrefix = "arXiv",
    primaryClass = "hep-th",
    doi = "10.1007/JHEP02(2026)157",
    journal = "JHEP",
    volume = "02",
    pages = "157",
    year = "2026"
}

@article{Anempodistov:2026dhi,
    author = "Anempodistov, Prokopii",
    title = "{Holographic two-point functions of heavy operators revisited}",
    eprint = "2603.28880",
    archivePrefix = "arXiv",
    primaryClass = "hep-th",
    month = "3",
    year = "2026",
    journal       = {}
}

@article{Berenstein:2004kk,
    author = "Berenstein, David",
    title = "{A Toy model for the AdS / CFT correspondence}",
    eprint = "hep-th/0403110",
    archivePrefix = "arXiv",
    doi = "10.1088/1126-6708/2004/07/018",
    journal = "JHEP",
    volume = "07",
    pages = "018",
    year = "2004"
}

@article{Berenstein:2020jen,
    author = "Berenstein, David and Holguin, Adolfo",
    title = "{Open giant magnons on LLM geometries}",
    eprint = "2010.02236",
    archivePrefix = "arXiv",
    primaryClass = "hep-th",
    doi = "10.1007/JHEP01(2021)080",
    journal = "JHEP",
    volume = "01",
    pages = "080",
    year = "2021"
}

@article{Skenderis:2007yb,
    author = "Skenderis, Kostas and Taylor, Marika",
    title = "{Anatomy of bubbling solutions}",
    eprint = "0706.0216",
    archivePrefix = "arXiv",
    primaryClass = "hep-th",
    reportNumber = "ITFA-2007-17",
    doi = "10.1088/1126-6708/2007/09/019",
    journal = "JHEP",
    volume = "09",
    pages = "019",
    year = "2007"
}

@article{Corley:2001zk,
    author = "Corley, Steve and Jevicki, Antal and Ramgoolam, Sanjaye",
    title = "{Exact correlators of giant gravitons from dual N=4 SYM theory}",
    eprint = "hep-th/0111222",
    archivePrefix = "arXiv",
    reportNumber = "BROWN-HET-1292",
    doi = "10.4310/ATMP.2001.v5.n4.a6",
    journal = "Adv. Theor. Math. Phys.",
    volume = "5",
    pages = "809--839",
    year = "2002"
}

@article{Berenstein:2022srd,
    author = "Berenstein, David and Wang, Shannon",
    title = "{BPS coherent states and localization}",
    eprint = "2203.15820",
    archivePrefix = "arXiv",
    primaryClass = "hep-th",
    doi = "10.1007/JHEP08(2022)164",
    journal = "JHEP",
    volume = "08",
    pages = "164",
    year = "2022"
}

@article{Turton:2024afd,
    author = "Turton, David and Tyukov, Alexander",
    title = "{Four-point correlators in $ \mathcal{N} $ = 4 SYM from AdS$_{5}$ bubbling geometries}",
    eprint = "2408.16834",
    archivePrefix = "arXiv",
    primaryClass = "hep-th",
    doi = "10.1007/JHEP10(2024)244",
    journal = "JHEP",
    volume = "10",
    pages = "244",
    year = "2024"
}

@article{Berenstein:2023vtd,
    author = "Berenstein, David and Maderazo, Elliot and Mancilla, Robinson and Ramirez, Anayeli",
    title = "{Chaotic LLM billiards}",
    eprint = "2305.19321",
    archivePrefix = "arXiv",
    primaryClass = "hep-th",
    doi = "10.1007/JHEP08(2024)056",
    journal = "JHEP",
    volume = "08",
    pages = "056",
    year = "2024"
}

@inproceedings{Zabrodin:2004cc,
    author = "Zabrodin, A.",
    title = "{Matrix models and growth processes: From viscous flows to the quantum Hall effect}",
    booktitle = "{NATO Advanced Study Institute: Marie Curie Training Course: Applications of Random Matrices in Physics}",
    eprint = "hep-th/0412219",
    archivePrefix = "arXiv",
    reportNumber = "ITEP-TH-80-04",
    pages = "261--318",
    month = "12",
    year = "2004"
}

@article{Douglas:1989ve,
    author = "Douglas, Michael R. and Shenker, Stephen H.",
    editor = "Brezin, E. and Wadia, S. R.",
    title = "{Strings in Less Than One-Dimension}",
    reportNumber = "RU-89-34",
    doi = "10.1016/0550-3213(90)90522-F",
    journal = "Nucl. Phys. B",
    volume = "335",
    pages = "635",
    year = "1990"
}

@article{Berenstein:2002jq,
    author = "Berenstein, David Eliecer and Maldacena, Juan Martin and Nastase, Horatiu Stefan",
    title = "{Strings in flat space and pp waves from N=4 superYang-Mills}",
    eprint = "hep-th/0202021",
    archivePrefix = "arXiv",
    doi = "10.1088/1126-6708/2002/04/013",
    journal = "JHEP",
    volume = "04",
    pages = "013",
    year = "2002"
}

@article{Beisert:2010jr,
    author = "Beisert, Niklas and others",
    title = "{Review of AdS/CFT Integrability: An Overview}",
    eprint = "1012.3982",
    archivePrefix = "arXiv",
    primaryClass = "hep-th",
    reportNumber = "AEI-2010-175, CERN-PH-TH-2010-306, HU-EP-10-87, HU-MATH-2010-22, KCL-MTH-10-10, UMTG-270, UUITP-41-10",
    doi = "10.1007/s11005-011-0529-2",
    journal = "Lett. Math. Phys.",
    volume = "99",
    pages = "3--32",
    year = "2012"
}

@article{Kazakov:2018ugh,
    author = "Kazakov, Vladimir",
    editor = "Ge, Mo-Lin and Niemi, Antti J. and Phua, Kok Khoo and Takhtajan, Leon A.",
    title = "{Quantum Spectral Curve of $\gamma$-twisted ${\cal N}=4$ SYM theory and fishnet CFT}",
    eprint = "1802.02160",
    archivePrefix = "arXiv",
    primaryClass = "hep-th",
    reportNumber = "LPTENS-18-02, LPTENS-18/02",
    doi = "10.1142/9789813233867_0016",
    journal = "Rev. Math. Phys.",
    pages = "293--342",
    year = "2018"
}

@article{Basso:2013vsa,
    author = "Basso, Benjamin and Sever, Amit and Vieira, Pedro",
    title = "{Spacetime and Flux Tube S-Matrices at Finite Coupling for N=4 Supersymmetric Yang-Mills Theory}",
    eprint = "1303.1396",
    archivePrefix = "arXiv",
    primaryClass = "hep-th",
    doi = "10.1103/PhysRevLett.111.091602",
    journal = "Phys. Rev. Lett.",
    volume = "111",
    number = "9",
    pages = "091602",
    year = "2013"
}

@article{Basso:2015zoa,
    author = "Basso, Benjamin and Komatsu, Shota and Vieira, Pedro",
    title = "{Structure Constants and Integrable Bootstrap in Planar N=4 SYM Theory}",
    eprint = "1505.06745",
    archivePrefix = "arXiv",
    primaryClass = "hep-th",
    month = "5",
    year = "2015",
    journal = {}
}

@article{Gromov:2013pga,
    author = "Gromov, Nikolay and Kazakov, Vladimir and Leurent, Sebastien and Volin, Dmytro",
    title = "{Quantum Spectral Curve for Planar $\mathcal{N} = 4$ Super-Yang-Mills Theory}",
    eprint = "1305.1939",
    archivePrefix = "arXiv",
    primaryClass = "hep-th",
    reportNumber = "IMPERIAL-TP-13-SL-02",
    doi = "10.1103/PhysRevLett.112.011602",
    journal = "Phys. Rev. Lett.",
    volume = "112",
    number = "1",
    pages = "011602",
    year = "2014"
}

@article{Gromov:2014caa,
    author = "Gromov, Nikolay and Kazakov, Vladimir and Leurent, S{\'e}bastien and Volin, Dmytro",
    title = "{Quantum spectral curve for arbitrary state/operator in AdS$_{5}$/CFT$_{4}$}",
    eprint = "1405.4857",
    archivePrefix = "arXiv",
    primaryClass = "hep-th",
    reportNumber = "NORDITA-2014-60-TCDMATH-14-05",
    doi = "10.1007/JHEP09(2015)187",
    journal = "JHEP",
    volume = "09",
    pages = "187",
    year = "2015"
}

@article{Takayama:2005yq,
    author = "Takayama, Yastoshi and Tsuchiya, Asato",
    title = "{Complex matrix model and fermion phase space for bubbling AdS geometries}",
    eprint = "hep-th/0507070",
    archivePrefix = "arXiv",
    reportNumber = "OU-HET-535",
    doi = "10.1088/1126-6708/2005/10/004",
    journal = "JHEP",
    volume = "10",
    pages = "004",
    year = "2005"
}

@article{Anempodistov:2025maj,
    author = "Anempodistov, Prokopii and Holguin, Adolfo and Kazakov, Vladimir and Murali, Harish",
    title = "{(Un)solvable matrix models for bps correlators}",
    eprint = "2508.20094",
    archivePrefix = "arXiv",
    primaryClass = "hep-th",
    doi = "10.1007/JHEP04(2026)069",
    journal = "JHEP",
    volume = "04",
    pages = "069",
    year = "2026"
}

@article{Aprile:2025hlt,
    author = "Aprile, Francesco and Giusto, Stefano and Russo, Rodolfo",
    title = "{Four-point correlators with BPS bound states in AdS$_{3}$ and AdS$_{5}$}",
    eprint = "2503.02855",
    archivePrefix = "arXiv",
    primaryClass = "hep-th",
    doi = "10.1007/JHEP08(2025)193",
    journal = "JHEP",
    volume = "08",
    pages = "193",
    year = "2025"
}

@article{Chen:2007du,
    author = "Chen, Bin and Cremonini, Sera and Donos, Aristomenis and Lin, Feng-Li and Lin, Hai and Liu, James T. and Vaman, Diana and Wen, Wen-Yu",
    editor = "Mueller, Bernt and Rotondo, Mary Ann and Tan, Chung-I",
    title = "{Bubbling AdS and droplet descriptions of BPS geometries in IIB supergravity}",
    eprint = "0704.2233",
    archivePrefix = "arXiv",
    primaryClass = "hep-th",
    reportNumber = "BROWN-HET-1480, MCTP-07-15, NSF-KITP-07-58, PARIS-2007-07",
    doi = "10.1088/1126-6708/2007/10/003",
    journal = "JHEP",
    volume = "10",
    pages = "003",
    year = "2007"
}

@article{Kazakov:1985ds,
    author = "Kazakov, V. A.",
    title = "{Bilocal Regularization of Models of Random Surfaces}",
    doi = "10.1016/0370-2693(85)91011-1",
    journal = "Phys. Lett. B",
    volume = "150",
    pages = "282--284",
    year = "1985"
}

@article{Kazakov:1985ea,
    author = "Kazakov, V. A. and Migdal, Alexander A. and Kostov, I. K.",
    title = "{Critical Properties of Randomly Triangulated Planar Random Surfaces}",
    doi = "10.1016/0370-2693(85)90669-0",
    journal = "Phys. Lett. B",
    volume = "157",
    pages = "295--300",
    year = "1985"
}

@article{David:1984tx,
    author = "David, F.",
    editor = "Brezin, E. and Wadia, S. R.",
    title = "{Planar Diagrams, Two-Dimensional Lattice Gravity and Surface Models}",
    reportNumber = "SACLAY-SPH-T-84-125",
    doi = "10.1016/0550-3213(85)90335-9",
    journal = "Nucl. Phys. B",
    volume = "257",
    pages = "45",
    year = "1985"
}

@article{Donos:2006iy,
    author = "Donos, Aristomenis",
    title = "{A Description of 1/4 BPS configurations in minimal type IIB SUGRA}",
    eprint = "hep-th/0606199",
    archivePrefix = "arXiv",
    reportNumber = "BROWN-HET-1470",
    doi = "10.1103/PhysRevD.75.025010",
    journal = "Phys. Rev. D",
    volume = "75",
    pages = "025010",
    year = "2007"
}

@article{Lunin:2008tf,
    author = "Lunin, Oleg",
    title = "{Brane webs and 1/4-BPS geometries}",
    eprint = "0802.0735",
    archivePrefix = "arXiv",
    primaryClass = "hep-th",
    reportNumber = "EFI-08-02",
    doi = "10.1088/1126-6708/2008/09/028",
    journal = "JHEP",
    volume = "09",
    pages = "028",
    year = "2008"
}

@article{Gava:2006pu,
    author = "Gava, Edi and Milanesi, Giuseppe and Narain, K. S. and O'Loughlin, Martin",
    title = "{1/8 BPS states in Ads/CFT}",
    eprint = "hep-th/0611065",
    archivePrefix = "arXiv",
    reportNumber = "SISSA-63-2006-EP, IC-2006-116",
    doi = "10.1088/1126-6708/2007/05/030",
    journal = "JHEP",
    volume = "05",
    pages = "030",
    year = "2007"
}

@article{Maldacena:1997re,
    author = "Maldacena, Juan Martin",
    title = "{The Large $N$ limit of superconformal field theories and supergravity}",
    eprint = "hep-th/9711200",
    archivePrefix = "arXiv",
    reportNumber = "HUTP-97-A097, HUTP-98-A097",
    doi = "10.4310/ATMP.1998.v2.n2.a1",
    journal = "Adv. Theor. Math. Phys.",
    volume = "2",
    pages = "231--252",
    year = "1998"
}

@article{Gubser:1998bc,
    author = "Gubser, S. S. and Klebanov, Igor R. and Polyakov, Alexander M.",
    title = "{Gauge theory correlators from noncritical string theory}",
    eprint = "hep-th/9802109",
    archivePrefix = "arXiv",
    reportNumber = "PUPT-1767",
    doi = "10.1016/S0370-2693(98)00377-3",
    journal = "Phys. Lett. B",
    volume = "428",
    pages = "105--114",
    year = "1998"
}

@article{Witten:1998qj,
    author = "Witten, Edward",
    title = "{Anti de Sitter space and holography}",
    eprint = "hep-th/9802150",
    archivePrefix = "arXiv",
    reportNumber = "IASSNS-HEP-98-15",
    doi = "10.4310/ATMP.1998.v2.n2.a2",
    journal = "Adv. Theor. Math. Phys.",
    volume = "2",
    pages = "253--291",
    year = "1998"
}

@article{Gromov:2017blm,
    author = "Gromov, Nikolay",
    title = "{Introduction to the Spectrum of $N=4$ SYM and the Quantum Spectral Curve}",
    eprint = "1708.03648",
    archivePrefix = "arXiv",
    primaryClass = "hep-th",
    month = "8",
    year = "2017",
    journal       = {}
}

@article{Daul:1993bg,
    author = "Daul, J. M. and Kazakov, V. A. and Kostov, I. K.",
    title = "{Rational theories of 2-D gravity from the two matrix model}",
    eprint = "hep-th/9303093",
    archivePrefix = "arXiv",
    reportNumber = "CERN-TH-6834-93",
    doi = "10.1016/0550-3213(93)90582-A",
    journal = "Nucl. Phys. B",
    volume = "409",
    pages = "311--338",
    year = "1993"
}


\makeatletter
\setcounter{section}{0}
\setcounter{subsection}{0}
\setcounter{equation}{0}
\renewcommand{\thesection}{\arabic{section}}
\renewcommand{\thesubsection}{\thesection.\arabic{subsection}}
\renewcommand{\theequation}{S\arabic{section}.\arabic{equation}}
\let\@sectioncntformat\@seccntformat
\def\@hangfrom@section#1#2#3{\@hangfrom{#1#2}\MakeTextUppercase{#3}}
\makeatother
\begin{center}
{\bfseries Supplementary Material}
\end{center}

\section{Derivation of universal cusp metric}
Let us study the functions that define LLM metric.

We assume that, in a neighbourhood  of the origin, the boundary of the droplet
has a multicritical cusp of the form
\[
x_1=-t+o(t),
\qquad
x_2=\pm t^\alpha+o(t^\alpha),
\qquad t\to0^+,
\]
with
$
1<\alpha<3,
\qquad
\alpha\neq2.
$

We now study the scaling limit of our functions under
$
(x_1,x_2,y)\mapsto
(\epsilon x_1,\epsilon x_2,\epsilon y)
$ as $
\epsilon\to 0^+ .
$

Then
\[
V(\epsilon x_1,\epsilon x_2,\epsilon y)
=
\int_D \frac{d^2z'}{\pi}\,
\frac{\epsilon z-z'}
{\left(|\epsilon z-z'|^2+\epsilon^2 y^2\right)^2},
\]
\[
\zeta(\epsilon x_1,\epsilon x_2,\epsilon y)-\frac12
=
-
\int_D \frac{d^2z'}{\pi}\,
\frac{\epsilon^2 y^2}
{\left(|\epsilon z-z'|^2+\epsilon^2 y^2\right)^2}.
\]
We now blow up the integration variables in such a way that the cut approaches
the origin from the left along the positive \(t\)-direction. Namely, we set
\[
x_1'=-\epsilon t,
\qquad
x_2'=\epsilon s,
\qquad
z'=\epsilon(-t+i s).
\]
And define
\[
D_\epsilon
:=
\left\{
(t,s)
\,:\,
\bigl(-\epsilon t,\epsilon s\bigr)\in D
\right\}.
\]
Since
$
d^2z'=dx_1'dx_2'
=
\epsilon^2\,dt\,ds,
$
we obtain\small{
\[
V(\epsilon x_1,\epsilon x_2,\epsilon y)
=
\frac1\epsilon
\int_{D_\epsilon}\frac{dt\,ds}{\pi}\,
\frac{x_1+t+i(x_2-s)}
{\left((x_1+t)^2+(x_2-s)^2+y^2\right)^2}.
\]
\[
\zeta(\epsilon x_1,\epsilon x_2,\epsilon y)-\frac12
=
-
\int_{D_\epsilon}\frac{dt\,ds}{\pi}\,
\frac{y^2}
{\left((x_1+t)^2+(x_2-s)^2+y^2\right)^2}.
\]}
We denote the cusp part of the droplet by
\[
D_{\mathrm{cusp}}
=
\left\{
(x_1',x_2'):\;
-\epsilon\Lambda<x_1'<0,\quad
h_-(x_1')<x_2'<h_+(x_1')
\right\}.
\]
In the blown-up variables
this becomes
\[
D_{\mathrm{cusp}}
=
\left\{
(t,s):\;
0<t<\Lambda,\quad
\frac{h_-(-\epsilon t)}{\epsilon}<s<
\frac{h_+(-\epsilon t)}{\epsilon}
\right\}.
\]

We first compute the contribution of \(D_{\mathrm{cusp}}\) to the scaling
limit of \(V\) and \(\zeta-\frac12\). These become
\[
V_{\mathrm{cusp}}
=
\frac1\epsilon
\int_{0}^{\Lambda} dt
\int_{h_-(-\epsilon t)/\epsilon}^{h_+(-\epsilon t)/\epsilon} ds\,
\frac{1}{\pi}\,
\frac{x_1+t+i(x_2-s)}
{\left[
(x_1+t)^2+(x_2-s)^2+y^2
\right]^2},
\]
and
\fontsize{8}{9}\selectfont{
\[
\zeta_{\mathrm{cusp}}-\frac12=-
\int_{0}^{\Lambda} dt
\int_{h_-(-\epsilon t)/\epsilon}^{h_+(-\epsilon t)/\epsilon} ds\,
\frac{1}{\pi}\,
\frac{y^2}
{\left[
(x_1+t)^2+(x_2-s)^2+y^2
\right]^2}.
\]}
Let
\begin{align*}
F_V(t,s)
=
\frac{x_1+t+i(x_2-s)}
{\left[(x_1+t)^2+(x_2-s)^2+y^2\right]^2},
\\
F_\zeta(t,s)
=
\frac{y^2}
{\left[(x_1+t)^2+(x_2-s)^2+y^2\right]^2}.
\end{align*}
For fixed finite \(\Lambda\), the denominator is positive on the region under
consideration, hence both integrands are differentiable with respect to \(s\).
Therefore Taylor's formula with the remainder in Lagrange form gives
\[
F(t,s)
=
F(t,0)
+
s\,\partial_s F(t,\xi),
\qquad
\xi \ \text{lies between }0\text{ and }s .
\]
This applies to \(F_\zeta\), and componentwise to the real and imaginary parts
of \(F_V\).

Let us first analyze the behavior of the first term.
The
\(s\)-integration then gives only the width of the cusp:
\[
\int_{h_-(-\epsilon t)/\epsilon}^{h_+(-\epsilon t)/\epsilon}
ds\,F(t,0)
=
F(t,0)\,
\frac{h_+(-\epsilon t)-h_-(-\epsilon t)}{\epsilon}.
\]

Using the cusp asymptotics,
\[
\frac{h_\pm(-\epsilon t)}{\epsilon}
=
\pm \epsilon^{\alpha-1}t^\alpha
+
o(\epsilon^{\alpha-1}),
\]
We obtain ($\Lambda$ is fixed here)
\[
\int_0^\Lambda dt
\int_{h_-(-\epsilon t)/\epsilon}^{h_+(-\epsilon t)/\epsilon}
ds\,F(t,s)
=
2\epsilon^{\alpha-1}
\int_0^\Lambda dt\,t^\alpha F(t,0)
+
o(\epsilon^{\alpha-1}).
\]

Secondly, there is a constant \(C>0\) such that
$
\left|\partial_s F(t,\xi)\right|\leq C
$
on the whole integration region. Hence the contribution of the second term is
bounded by
\[
\left|
\int_0^\Lambda dt
\int_{h_-(-\epsilon t)/\epsilon}^{h_+(-\epsilon t)/\epsilon}
ds\,
s\,\partial_sF(t,\xi)
\right|
\leq
C
\int_0^\Lambda dt
\int_{h_-(-\epsilon t)/\epsilon}^{h_+(-\epsilon t)/\epsilon}
|s|\,ds .
\]
We have, for sufficiently small \(\epsilon\),
\[
|s|\leq C \epsilon^{\alpha-1}t^\alpha,
\qquad
\frac{h_+(-\epsilon t)-h_-(-\epsilon t)}{\epsilon}
\leq
C \epsilon^{\alpha-1}t^\alpha .
\]
Hence
\[
\left|
\int_0^\Lambda dt
\int_{h_-(-\epsilon t)/\epsilon}^{h_+(-\epsilon t)/\epsilon}
ds\,
s\,\partial_sF(t,\xi)
\right|
\leq
C
\epsilon^{2\alpha-2}
\int_0^\Lambda t^{2\alpha}\,dt
=
O(
\epsilon^{2\alpha-2}
).
\]
Thus, for fixed finite \(\Lambda\), the second term in Taylor's formula gives a
subleading contribution. 

Applying this to \(F_V\) and \(F_\zeta\), we finally get
\[
V_{\mathrm{cusp}}
=
\frac{2\epsilon^{\alpha-2}}{\pi}
\int_0^\Lambda dt\,
t^\alpha
\frac{x_1+t+i x_2}
{\left[(x_1+t)^2+x_2^2+y^2\right]^2}
+
o(\epsilon^{\alpha-2}),
\]
and
\[
\zeta_{\mathrm{cusp}}-\frac12
=
-\frac{2\epsilon^{\alpha-1}}{\pi}
\int_0^\Lambda dt\,
t^\alpha
\frac{y^2}
{\left[(x_1+t)^2+x_2^2+y^2\right]^2}
+
o(\epsilon^{\alpha-1}) .
\]

Firstly, we study the case \(1<\alpha<2\).
For fixed finite \(\Lambda\), the cusp contributions are written above. It remains to estimate the complementary region
$
D_{\mathrm{out}}=D\setminus D_{\mathrm{cusp}}.
$

On \(D_{\mathrm{out}}\) we have
\[
\left|
\frac{\epsilon z-z'}
{\bigl(|\epsilon z-z'|^2+\epsilon^2y^2\bigr)^2}
\right|
\le
\frac{C}{|z'|^3},
\]
and
\[
\left|
\frac{\epsilon^2y^2}
{\bigl(|\epsilon z-z'|^2+\epsilon^2y^2\bigr)^2}
\right|
\le
C\epsilon^2\frac1{|z'|^4}.
\]

We can write
\begin{align*}
|V_{\mathrm{out}}|
\le&
C\int^{-\epsilon\Lambda}_{-L}dx_1'
\int_{-Cx_1^\alpha}^{Cx_1^\alpha}dx_2'\,\frac1{(-x_1')^3}
\le
C\int^{-\epsilon\Lambda}_{-L}dx'_1\,(-x_1')^{\alpha-3}=\\&=C((\epsilon\Lambda)^{\alpha-2}-L^{\alpha-2})\le
C\epsilon^{\alpha-2}\Lambda^{\alpha-2}.
\end{align*}
Similarly,
\begin{align*}
&\left|
\zeta_{\mathrm{out}}-\frac12
\right|
\le
C\epsilon^2
\int^{-\epsilon\Lambda}_{-L}dx_1'
\int_{-Cx_1'^\alpha}^{Cx_1^\alpha}dx_2\,\frac1{(-x_1')^4}\le
\\&\le
C\epsilon^2\int^{-\epsilon\Lambda}_{-L}dx'_1\,(-x_1'^{\alpha-4}=C\epsilon^2((\epsilon\Lambda)^{\alpha-3}-L^{\alpha-3})\epsilon^2\le
C\epsilon^{\alpha-1}\Lambda^{\alpha-3}.
\end{align*}
Dividing by the corresponding powers of \(\epsilon\) and taking the limit \(\epsilon\to0\) and then \(\Lambda\to\infty\), we obtain
\begin{align*}
&\epsilon^{2-\alpha}
V(\epsilon x_1,\epsilon x_2,\epsilon y)=\\
&=
\frac{2}{\pi}
\int_0^\Lambda dt\,t^\alpha
\frac{x_1+t+i x_2}
{\bigl[(x_1+t)^2+x_2^2+y^2\bigr]^2}
+
O(\Lambda^{\alpha-2}),
\end{align*}
and
\begin{align*}
&\epsilon^{1-\alpha}
\left(
\zeta(\epsilon x_1,\epsilon x_2,\epsilon y)-\frac12
\right)
=\\
&=-\frac{2}{\pi}
\int_0^\Lambda dt\,t^\alpha
\frac{y^2}
{\bigl[(x_1+t)^2+x_2^2+y^2\bigr]^2}
+
O(\Lambda^{\alpha-3}).
\end{align*}
 Both
integrals converge, while the correction terms vanish.
Therefore, sending \(\Lambda\to\infty\) and introducing spherical coordinates
\[
x_1=r\cos\theta,
\qquad
x_2=r\sin\theta\cos\varphi,
\qquad
y=r\sin\theta\sin\varphi .
\] 
and function 
\[
S_\alpha(\theta)=
-\frac{\sin\bigl((\alpha-1)\theta\bigr)}
{\sin(\pi\alpha)\sin\theta}.
\]
we get
\[
V(\epsilon x_1,\epsilon x_2,\epsilon y)
=
\epsilon^{\alpha-2}r^{\alpha-2}
\left[
\alpha S_\alpha(\theta)
+
i\cos\varphi\,S_\alpha'(\theta)
\right]
\].
and
\begin{align*}
\zeta(\epsilon x_1,\epsilon x_2,\epsilon y)-\frac12
=
-\epsilon^{\alpha-1}r^{\alpha-1}
\sin\theta\,\sin^2\varphi\,
S_\alpha'(\theta).
\end{align*}
Now we study the case
\(2<\alpha<3\).

The estimate for \(\zeta\) is unchanged from the previous case, but 
the behavior of \(V=V_{\mathrm{out}}+V_{\mathrm{cusp}}\) is different. The cusp contribution $V_{\mathrm{cusp}}=O(\epsilon^{\alpha-2})$.

 Thus the leading term of \(V\) is not produced by the local cusp scaling region.
Indeed, the limiting kernel
$-\frac{z'}{|z'|^4}$
is integrable on the droplet when $\alpha>2$. 
And the whole integral converges to
\[V(\epsilon x_1,\epsilon x_2,\epsilon y)=-\int_D\frac{d^2z'}{\pi}\frac{z'}{|z'|^4}+O(\epsilon^{\alpha-2}).\]Thus \(V\) is finite and its leading coefficient depends on the whole droplet \(D\), not only on the local cusp geometry. 

Finally, consider  the case \(\alpha>3\). In this case even \(\zeta\) is no longer governed by the local cusp scaling. Indeed, $\frac1{|z'|^4}$ is integrable on the droplet when $\alpha>3$.
The whole integral once again converges to:\[\zeta(\epsilon x_1,\epsilon x_2,\epsilon y)-\frac12=-\epsilon^2\frac{y^2}{\pi}\int_D\frac{d^2z'}{|z'|^4}+o(\epsilon^2).\]
The leading coefficient is therefore non-universal and depends on the whole droplet. Thus for \(\alpha>3\) the local cusp scaling is subleading and does not define a critical local metric; there is no need to analyze \(V\) further for the critical geometry.

\vspace{0.5cm}
\section{Naked singularity}

Now let us calculate the Kretschmann scalar for the scaled metric. Denoting
\begin{equation}
    \begin{aligned}
\mathcal{H}_{\alpha} \equiv \mathcal{H}(r,\theta),\qquad
\mathcal{H}_r \equiv \partial_r \mathcal{H},\qquad
\mathcal{H}_\theta \equiv \partial_\theta \mathcal{H},\qquad\\
\mathcal{H}_{rr} \equiv \partial_r^2 \mathcal{H},\qquad
\mathcal{H}_{\theta\theta} \equiv \partial_\theta^2 \mathcal{H},\qquad
\mathcal{H}_{r\theta} \equiv \partial_r \partial_\theta \mathcal{H} ,\notag
\end{aligned}
\end{equation}

one obtains
\begin{equation}
    \begin{aligned}
\mathcal{K} =  \frac{1}{\mathcal{H}^2 r^4}\Bigg[
9 \mathcal{H}^2 \mathcal{H}_{\theta\theta}^2
+2 \mathcal{H}^2 \mathcal{H}_{rr} \mathcal{H}_{\theta\theta} r^2
-10 \mathcal{H} \mathcal{H}_r^2 \mathcal{H}_{\theta\theta} r^2+\\
+26 \mathcal{H}^2 \mathcal{H}_r \mathcal{H}_{\theta\theta} r
+8 \mathcal{H}^2 \mathcal{H}_\theta \mathcal{H}_{\theta\theta} \cot\theta
-10 \mathcal{H} \mathcal{H}_\theta^2 \mathcal{H}_{\theta\theta}+\\
+16 \mathcal{H}^2 \mathcal{H}_{r\theta}^2 r^2
-32 \mathcal{H}^2 \mathcal{H}_\theta \mathcal{H}_{r\theta} r
+9 \mathcal{H}^2 \mathcal{H}_{rr}^2 r^4-\\
-10 \mathcal{H} \mathcal{H}_r^2 \mathcal{H}_{rr} r^4
+10 \mathcal{H}^2 \mathcal{H}_r \mathcal{H}_{rr} r^3
+8 \mathcal{H}^2 \mathcal{H}_\theta \mathcal{H}_{rr} r^2 \cot\theta-\\
-10 \mathcal{H} \mathcal{H}_\theta^2 \mathcal{H}_{rr} r^2
+14 \mathcal{H}_\theta^4
-40 \mathcal{H} \mathcal{H}_\theta^3 \cot\theta-\\
-16 \mathcal{H}^2 \mathcal{H}_\theta^2
\left(
-2\cot^2\theta-\csc^2\theta
\right)-\\
-50 \mathcal{H} \mathcal{H}_\theta^2 \mathcal{H}_r r
+28 \mathcal{H}_\theta^2 \mathcal{H}_r^2 r^2
-40 \mathcal{H} \mathcal{H}_\theta \mathcal{H}_r^2 r^2 \cot\theta+\\
+104 \mathcal{H}^2 \mathcal{H}_\theta \mathcal{H}_r r \cot\theta
+14 \mathcal{H}_r^4 r^4
-50 \mathcal{H} \mathcal{H}_r^3 r^3
+65 \mathcal{H}^2 \mathcal{H}_r^2 r^2
\Bigg].\notag
\end{aligned}
\end{equation}

Here 
\begin{equation}
    \mathcal{H}(r,\theta) \equiv r^{\frac{3-\alpha}{2}}\left[\frac{
-2\sin^{3}\theta\,\sin(\pi\alpha)}{\left((\alpha-2)\sin(\alpha\theta)
+\alpha\sin((2-\alpha)\theta)\right)}
\right]^{1/2}.\notag
\end{equation}
Evaluating this expression and expanding it around $\theta = \pm \pi$, one obtains asymptotic form 
$\mathcal{K}
\sim\frac{657}{16}r^{-(\alpha+1)}
(\pi\mp\theta)^{-1}$.

\section{D3-brane interpretation of the cusp metric}
The metric for the stack of D3-branes is
\begin{equation}
    ds^2
    =
    H_{\rm D3}^{-1/2}ds_{\mathbb R^{1,3}}^2
    +
    H_{\rm D3}^{1/2}
    \left(
    dr^2+r^2d\theta^2+r^2\sin^2\theta\,d\Omega_4^2
    \right),\notag
\end{equation}
where $H_{D3}$ is a harmonic function in six-dimensional flat space transversal to the branes

Let us consider a D3-brane distribution in a two-dimensional plane of the
transverse $\mathbb R^6$, with coordinates $(X,Y)$. The remaining four
transverse directions are denoted collectively by $\vec u\in\mathbb R^4$.
We take the cusp to point to the left of the origin. Thus, locally, the
support of the distribution is
\begin{equation}
    X<0,\qquad |Y|<C(-X)^\alpha ,\notag
\end{equation}
where we assume
\begin{equation}
    1<\alpha<2 .\notag
\end{equation}
The upper bound is not essential for the line-source integral below, but it is
the range relevant for the usual cusp interpretation.

We now zoom into the cusp by the isotropic scaling
\begin{equation}
    X=\epsilon x,\qquad Y=\epsilon y .\notag
\end{equation}
In the scaled coordinates the support becomes
\begin{equation}
    x<0,\qquad |y|<C\epsilon^{\alpha-1}(-x)^\alpha .\notag
\end{equation}
Therefore, in the limit $\epsilon\to0$, the support collapses to the left
semi-axis,
\begin{equation}
    x<0,\qquad y=0 .\notag
\end{equation}
To obtain a finite non-zero limiting density, we simultaneously scale the
two-dimensional density as
\begin{equation}
    \sigma_\epsilon(x,y)
    =
    \epsilon^{1-\alpha}\sigma_0\,
    \mathbf 1_{\{|y|<C\epsilon^{\alpha-1}(-x)^\alpha,\ x<0\}} .\notag
\end{equation}
Then, in the sense of distributions,
\begin{equation}
    \sigma_\epsilon(x,y)
    \longrightarrow
    \lambda(s)\delta(y),
    \qquad x=-s,\qquad s>0 ,\notag
\end{equation}
where the effective line density is
\begin{equation}
    \lambda(s)=\lambda_0 s^\alpha,
    \qquad 
    \lambda_0=2C\sigma_0 .\notag
\end{equation}
For a one-sided cusp rather than a symmetric cusp, the numerical factor $2C$
is simply replaced by the corresponding local width coefficient. In what
follows this overall normalization is absorbed into $\lambda_0$.

The D3-brane harmonic function is therefore
\begin{equation}
    H_{\rm D3}(x,R)
    =
    4\pi g_s(\alpha')^2
    \int_0^\infty
    \frac{ds\,\lambda_0 s^\alpha}
    {\left[(x+s)^2+R^2\right]^2},\notag
\end{equation}
where
\begin{equation}
    R^2=Y^2+|\vec u|^2\notag
\end{equation}
is the radial distance away from the limiting line source. It is useful to
define
\begin{equation}
    q_\alpha=4\pi g_s(\alpha')^2\lambda_0 .\notag
\end{equation}
Then
\begin{equation}
    H_{\rm D3}(x,R)
    =
    q_\alpha
    \int_0^\infty
    \frac{ds\,s^\alpha}
    {\left[(x+s)^2+R^2\right]^2}.\notag
\end{equation}

We now introduce polar coordinates adapted to the left semi-axis:
\begin{equation}
    x=r\cos\theta,\qquad R=r\sin\theta,
    \qquad 0<\theta<\pi .\notag
\end{equation}
Equivalently,
\begin{equation}
    dx^2+dR^2+R^2d\Omega_4^2
    =
    dr^2+r^2d\theta^2+r^2\sin^2\theta\,d\Omega_4^2 .\notag
\end{equation}
Changing variables $s=rt$, we find
\begin{equation}
    H_{\rm D3}(r,\theta)
    =
    q_\alpha r^{\alpha-3}
    I_\alpha(\theta),\notag
\end{equation}
where
\begin{equation}
    I_\alpha(\theta)
    =
    \int_0^\infty
    \frac{dt\,t^\alpha}
    {\left(t^2+2t\cos\theta+1\right)^2}.\notag
\end{equation}
For $1<\alpha<3$, this integral evaluates to
\begin{equation}
    I_\alpha(\theta)
    =
    -\frac{\pi}{4\sin(\pi\alpha)\sin^3\theta}
    \left[
    (\alpha-2)\sin(\alpha\theta)
    +
    \alpha\sin((2-\alpha)\theta)
    \right].\notag
\end{equation}
Therefore
\begin{equation}
    H_{\rm D3}(r,\theta)
    =
    -\frac{\pi q_\alpha}{4\sin(\pi\alpha)}
    r^{\alpha-3}
    \frac{
    (\alpha-2)\sin(\alpha\theta)
    +
    \alpha\sin((2-\alpha)\theta)
    }
    {\sin^3\theta}.\notag
\end{equation}

The critical cusp metric is written as
\begin{equation}
    ds_{\rm crit}^2
    =
    \mathcal H_\alpha(r,\theta)ds_{\mathbb R^{1,3}}^2
    +
    \mathcal H_\alpha(r,\theta)^{-1}
    \left(
    dr^2+r^2d\theta^2+r^2\sin^2\theta\,d\Omega_4^2
    \right).\notag
\end{equation}
Hence the quantity to compare with the universal critical warp factor is
\begin{equation}
    \mathcal H_\alpha=H_{\rm D3}^{-1/2}.\notag
\end{equation}
Using the expression above for $H_{\rm D3}$, we obtain
\begin{equation}
    H_{\rm D3}^{-1/2}
    =
    \left[
    -\frac{4\sin(\pi\alpha)}{\pi q_\alpha}
    \right]^{1/2}
    r^{\frac{3-\alpha}{2}}
    \left[
    \frac{\sin^3\theta}
    {
    (\alpha-2)\sin(\alpha\theta)
    +
    \alpha\sin((2-\alpha)\theta)
    }
    \right]^{1/2}.\notag
\end{equation}
The universal critical warp factor is
\begin{equation}
    \mathcal H_\alpha(r,\theta)
    =
    r^{\frac{3-\alpha}{2}}
    \left[
    \frac{
    -2\sin^3\theta\,\sin(\pi\alpha)}
    {
    (\alpha-2)\sin(\alpha\theta)
    +
    \alpha\sin((2-\alpha)\theta)
    }
    \right]^{1/2}.\notag
\end{equation}
Thus the radial dependence and the angular dependence agree exactly. The
remaining difference is only the overall normalization. Matching the two
expressions fixes
\begin{equation}
    q_\alpha=\frac{2}{\pi}.\notag
\end{equation}
Equivalently, in terms of the line-density normalization,
\begin{equation}
    \lambda_0
    =
    \frac{1}{2\pi^2 g_s(\alpha')^2}.\notag
\end{equation}

Therefore the scaling limit of a left-oriented $\alpha$-cusp distribution of
D3-branes produces a semi-infinite line distribution
\begin{equation}
    \lambda(s)\propto s^\alpha,\qquad s>0,\notag
\end{equation}
and the corresponding D3-brane metric is precisely of the critical form
\begin{equation}
    ds^2
    =
    \mathcal H_\alpha ds_{\mathbb R^{1,3}}^2
    +
    \mathcal H_\alpha^{-1}
    \left(
    dr^2+r^2d\theta^2+r^2\sin^2\theta\,d\Omega_4^2
    \right),\notag
\end{equation}
with
\begin{equation}
    \mathcal H_\alpha=H_{\rm D3}^{-1/2}.\notag
\end{equation}

As a useful check, for $\alpha=3/2$ one finds
\begin{equation}
    (\alpha-2)\sin(\alpha\theta)
    +
    \alpha\sin((2-\alpha)\theta)
    =
    -\frac12\sin\frac{3\theta}{2}
    +
    \frac32\sin\frac{\theta}{2}
    =
    2\sin^3\frac{\theta}{2}.\notag
\end{equation}
Hence
\begin{equation}
    \mathcal H_{3/2}(r,\theta)
    =
    2\sqrt2\,r^{3/4}\cos^{3/2}\frac{\theta}{2}, \notag
\end{equation}
which vanishes at $\theta=\pi$, i.e. precisely along the left semi-axis.

\end{document}